\documentclass[twocolumn,preprintnumbers,amsmath,amssymbm,prl]{revtex4}
\usepackage{epsfig}
\usepackage{graphicx}

\begin{document}
\title{Stationary Scalar Clouds Around Rotating Black Holes}
\author{Shahar Hod}
\affiliation{The Ruppin Academic Center, Emeq Hefer 40250, Israel}
\affiliation{ } \affiliation{The Hadassah Institute, Jerusalem
91010, Israel}
\date{\today}

\begin{abstract}
\ \ \ Motivated by novel results in the theory of wave dynamics in
black-hole spacetimes, we analyze the dynamics of a massive scalar
field surrounding a rapidly rotating Kerr black hole. In particular,
we report on the existence of stationary ({\it infinitely}
long-lived) regular field configurations in the background of
maximally rotating black holes. The effective height of these scalar
``clouds" above the central black hole is determined analytically.
Our results support the possible existence of stationary scalar
field dark matter distributions surrounding rapidly rotating black
holes.
\end{abstract}
\bigskip
\maketitle

{\bf I. Introduction.\ \ }
The `no-hair' conjecture \cite{Whee,Car}
has played a key role in the development of black-hole physics
\cite{BekTod1,BekTod2,BekTod3,Nun}. This conjecture asserts that
black holes are fundamental objects --- they should be described by
only three externally observable (conserved) parameters: mass,
charge, and angular momentum.

The physical idea behind the no-hair conjecture is based on an
intuitive (and sometimes oversimplified) picture according to which
all matter fields which are present in the exterior of a black hole
would eventually be swallowed by the black hole or be radiated away
to infinity (with the exception of fields which are associated with
conserved charges) \cite{BekTod2,Hod11}. In accord with this logic,
various no-hair theorems indeed exclude {\it static}
(time-independent) scalar fields \cite{Chas}, massive vector fields
\cite{BekVec}, and spinor fields \cite{Hart} from the exterior of
black holes.

It should be stressed, however, that the various no-hair theorems
\cite{Whee,Car,BekTod1,BekTod2,BekTod3,Nun,Hod11,Chas,BekVec,Hart}
do {\it not} rule out the existence of {\it time-dependent} field
configurations in the black-hole exterior. Indeed, in a very nice
work Barranco et. al. \cite{Barr} have recently found time-{\it
decaying} regular scalar field configurations surrounding a
Schwarzschild black hole that can survive for relatively long times
(as compared to the dynamical timescale set by the mass of the black
hole).

Ultra-light scalar fields have been invoked in recent years as
possible candidates to play the role of the dark matter component of
our universe (see \cite{Barr,Hu1,Hu2,Hu3} and references therein).
Given the fact that most galaxies seem to contain a super-massive
black hole at their centers \cite{Tal}, it was pointed out in
\cite{Barr} that in order to be a viable candidate for the dark
matter halo a scalar field configuration must be able to survive for
(at least) cosmological timescales. In this context, the important
conclusion presented in \cite{Barr} is very encouraging:
time-decaying configurations made of ultra-light scalar fields
surrounding a supermassive Schwarzschild black hole can indeed
survive for such extremely long periods.

The main aim of the present work is to extend the important results
of \cite{Barr} in three new directions:
\newline
(1) The analysis of \cite{Barr} was restricted to a {\it
spherically} symmetric Schwarzschild black-hole background. However,
it is well-known that realistic black holes generally rotate about
their axis and are therefore not spherical. Thus, an astrophysically
realistic model of wave dynamics in black-hole spacetimes must
involve a {\it non}-spherical background geometry with {\it angular
momentum} \cite{Hod2000}. In the present work we consider such
non-spherical backgrounds and analyze the dynamics of massive scalar
fields in realistic rotating (Kerr) black-hole spacetimes.
\newline
(2) The regular scalar configurations considered in \cite{Barr} have
long yet {\it finite} lifetimes. In the present work we shall prove
the existence of stationary (i.e. {\it infinitely} long-lived)
regular field configurations surrounding rotating black holes.
\newline
(3) The focus in \cite{Barr} was mainly on black hole--scalar field
configurations characterized by the dimensionless product $M\mu\ll
1$, where $M$ is the mass of the central black hole and $\mu$ is the
mass of the surrounding scalar field. In the present work we shall
analyze the complementary regime of $M\mu>{1\over 2}$.

{\bf II. Description of the system.\ \ } The physical system we
consider consists of a massive scalar field coupled to a rotating
Kerr black hole of mass $M$ and angular momentum per unit mass $a$.
In order to facilitate a fully {\it analytical} study, we shall
assume that the black hole is maximally rotating with $a=M$. Similar
results can be obtained (with the cost of a more involved analysis)
for non-extremal rotating black holes. In the present study we shall
restrict ourselves to the case of a test scalar field in the
background of the rotating Kerr black-hole spacetime. We shall
therefore use the terminology of \cite{Barr} and talk about scalar
``clouds" surrounding the black hole rather than a genuine scalar
``hair" \cite{Notenonlin}.

The dynamics of a massive scalar field $\Psi$ in the Kerr spacetime
is governed by the Klein-Gordon equation \cite{Teuk}
\begin{equation}\label{Eq1}
(\nabla^a \nabla_a -\mu^2)\Psi=0\  .
\end{equation}
(Here $\mu$ stands for ${\cal M}G/\hbar c$, where ${\cal M}$ is the
mass of the scalar field. We shall use units in which $G=c=\hbar=1$.
In these units $\mu$ has the dimensions of 1/length.) One may
decompose the field as
\begin{equation}\label{Eq2}
\Psi_{lm}(t,r,\theta,\phi)=e^{im\phi}S_{lm}(\theta;M\omega)R_{lm}(r;M,\omega)e^{-i\omega
t}\ ,
\end{equation}
where $(t,r,\theta,\phi)$ are the Boyer-Lindquist coordinates,
$\omega$ is the (conserved) frequency of the mode, $l$ is the
spheroidal harmonic index, and $m$ is the azimuthal harmonic index
with $-l\leq m\leq l$. (We shall henceforth omit the indices $l$ and
$m$ for brevity.) With the decomposition (\ref{Eq2}), $R$ and $S$
obey radial and angular equations both of confluent Heun type
coupled by a separation constant $K(M\omega)$
\cite{Heun,Fiz1,Teuk,Abram,Stro}.

The angular functions $S(\theta;a\omega)$ are the spheroidal
harmonics which are solutions of the angular equation
\cite{Heun,Fiz1,Teuk,Abram,Stro}
\begin{eqnarray}\label{Eq3}
{1\over {\sin\theta}}{\partial \over
{\partial\theta}}\Big(\sin\theta {{\partial
S}\over{\partial\theta}}\Big) +\Big[K+M^2(\mu^2-\omega^2) \nonumber
\\ -M^2(\mu^2-\omega^2)\cos^2\theta-{{m^2}\over{\sin^2\theta}}\Big]S=0\
.
\end{eqnarray}
The angular functions are required to be regular at the poles
$\theta=0$ and $\theta=\pi$. These boundary conditions pick out a
discrete set of eigenvalues $\{K_{lm}\}$ labeled by the integers $l$
and $m$. For $M^2(\mu^2-\omega^2)\lesssim m^2$ one can treat
$M^2(\omega^2-\mu^2)\cos^2\theta$ in Eq. (\ref{Eq3}) as a
perturbation term on the generalized Legendre equation and obtain
the perturbation expansion \cite{Abram}
\begin{equation}\label{Eq4}
K_{lm}+M^2(\mu^2-\omega^2)=l(l+1)+\sum_{k=1}^{\infty}c_kM^{2k}(\mu^2-\omega^2)^k\
\end{equation}
for the separation constants $K_{lm}$. The expansion coefficients
$\{c_k(l,m)\}$ are given in Ref. \cite{Abram}.

The radial Teukolsky equation is given by \cite{Teuk,Hodcen,Stro}
\begin{equation}\label{Eq5}
\Delta{{d} \over{dr}}\Big(\Delta{{dR}\over{dr}}\Big)+\Big[H^2
+\Delta[2mM\omega-K-\mu^2(r^2+M^2)]\Big]R=0\ ,
\end{equation}
where $\Delta\equiv (r-M)^2$ and $H\equiv (r^2+M^2)\omega-mM$. The
degenerate zero of $\Delta$, $r_H=M$, is the location of the
black-hole horizon.

We are interested in solutions of the radial equation (\ref{Eq5})
with the physical boundary conditions of purely ingoing waves at the
black-hole horizon (as measured by a comoving observer) and a
decaying (bounded) solution at spatial infinity
\cite{DamDer,Zour,Det,Dolan,HodSO,Hodst}. That is,
\begin{equation}\label{Eq6}
R \sim
\begin{cases}
{1\over r}e^{-\sqrt{\mu^2-\omega^2}y} & \text{ as }
r\rightarrow\infty\ \ (y\rightarrow \infty)\ ; \\
e^{-i (\omega-m\Omega)y} & \text{ as } r\rightarrow r_H\ \
(y\rightarrow -\infty)\ ,
\end{cases}
\end{equation}
where the ``tortoise" radial coordinate $y$ is defined by
$dy=[(r^2+M^2)/\Delta]dr$. Here $\Omega=1/2M$ is the angular
velocity of the black-hole horizon.

Note that a bound state (a state decaying exponentially at spatial
infinity) is characterized by $\omega^2<\mu^2$. For a given mass
parameter $\mu$, the boundary conditions (\ref{Eq6}) single out a
discrete set of resonances $\{\omega_n(\mu)\}$ which correspond to
the bound states of the massive field
\cite{DamDer,Zour,Det,Dolan,HodSO,Hodst}. Stationary resonances,
which are the solutions we are interested in in this paper, are
characterized by $\Im\omega=0$. (We note that, in addition to the
bound states of the massive field, the field also has an infinite
set of discrete quasinormal resonances
\cite{Will,ZhKo,Hodmassive,Massfr} which are characterized by
outgoing waves at spatial infinity.)

{\bf III. The stationary scalar resonances.\ \ } As we shall now
show, the field (\ref{Eq2}) with $\omega=m\Omega$ describes a {\it
stationary} regular solution of the wave equation (\ref{Eq1}). It is
convenient to define a new dimensionless variable
\begin{equation}\label{Eq7}
x\equiv {{r-r_H}\over {r_H}}\  ,
\end{equation}
in terms of which the radial equation (\ref{Eq5}) becomes
\begin{equation}\label{Eq8}
x^2{{d^2R}\over{dx^2}}+2x{{dR}\over{dx}}+VR=0\  ,
\end{equation}
where
$V\equiv(m^2/4-M^2\mu^2)x^2+(m^2-2M^2\mu^2)x+(-K_{lm}+2m^2-2M^2\mu^2)$.
Defining
\begin{equation}\label{Eq9}
f\equiv xR\ \ \ \text{and}\ \ \ z\equiv 2{\sqrt{\mu^2-m^2/4}}Mx\ ,
\end{equation}
one obtains the radial equation
\begin{equation}\label{Eq10}
{{d^2f}\over{dz^2}}+\Big[-{{1}\over{4}}+{{\kappa}\over{z}}+{{{1\over
4}-\beta^2}\over{z^2}}\Big]f=0\ ,
\end{equation}
with
\begin{equation}\label{Eq11}
\kappa\equiv {{m^2-2M^2\mu^2}\over{\sqrt{4M^2\mu^2-m^2}}}\ \ \
\text{and}\ \ \ \beta^2\equiv K_{lm}+{1\over 4}-2m^2+2M^2\mu^2\ .
\end{equation}
We shall assume without loss of generality that $\Re\beta\geq0$.
Equation (\ref{Eq10}) is the familiar Whittaker equation; its
solutions can be expressed in terms of the confluent hypergeometric
functions $M(a,b,z)$ \cite{Morse,Abram,HodSO}
\begin{equation}\label{Eq12}
R=Az^{-{1\over 2}+\beta}e^{-{1\over 2}z}M({1\over
2}+\beta-\kappa,1+2\beta,z)+B(\beta\to -\beta)\ ,
\end{equation}
where $A$ and $B$ are constants. The notation $(\beta\to -\beta)$
means ``replace $\beta$ by $-\beta$ in the preceding term."

The near-horizon ($z\to 0$) limit of Eq. (\ref{Eq12}) yields
\cite{Morse,Abram}
\begin{equation}\label{Eq13}
R\to Az^{-{1\over 2}+\beta}+Bz^{-{1\over 2}-\beta}\ .
\end{equation}
Regularity of the solution at the horizon ($R$ is bounded for $z\to
0$) requires
\begin{equation}\label{Eq14}
B=0\ \ \ \text{and}\ \ \ \Re\beta\geq {1\over 2}\  .
\end{equation}

Approximating Eq. (\ref{Eq12}) for $z\to\infty$ one gets
\cite{Morse,Abram}
\begin{eqnarray}\label{Eq15}
R\to A\Big[{{\Gamma(1+2\beta)}\over{\Gamma({1\over
2}+\beta-\kappa)}}z^{-1-\kappa}e^{{1\over 2}z} \nonumber
\\+{{\Gamma(1+2\beta)}\over{\Gamma({1\over
2}+\beta+\kappa)}}z^{-1+\kappa}(-1)^{-{1\over
2}-\beta+\kappa}e^{-{1\over 2}z}\Big]\ .
\end{eqnarray}
A bound state is characterized by a decaying field at spatial
infinity. The coefficient $1/\Gamma({1\over 2}+\beta-\kappa)$ of the
growing exponent $e^{{1\over 2}z}$ in Eq. (\ref{Eq15}) should
therefore vanish. Using the well-known pole structure of the Gamma
functions \cite{Abram}, we find the resonance condition for the
stationary bound-states of the field:
\begin{equation}\label{Eq16}
\kappa={1\over 2}+\beta+n  ,
\end{equation}
where the resonance parameter $n$ is a non-negative integer
($n=0,1,2,...$). Substituting Eqs. (\ref{Eq14}) and (\ref{Eq16})
into Eq. (\ref{Eq12}), one obtains the compact form
\begin{equation}\label{Eq17}
R=Az^{-{1\over 2}+\beta}e^{-{1\over 2}z}L^{(2\beta)}_n(z)\
\end{equation}
for the radial solutions, where $L^{(2\beta)}_n(z)$ are the
generalized Laguerre Polynomials (see Eq. $13.6.9$ of \cite{Abram}).

We note that the r.h.s of the resonance condition (\ref{Eq16}) is
positive definite. Taking cognizance of Eq. (\ref{Eq11}) with
$\kappa>0$, one concludes that the bound-state resonances must lie
within the band
\begin{equation}\label{Eq18}
{m\over 2}<M\mu<{m\over {\sqrt{2}}}\  .
\end{equation}
Note that the inequality (\ref{Eq18}) excludes the existence of
regular {\it static} solutions with $m=0$ ($\omega=0$). Of course,
this finding is in agreement with the well-known no-hair theorems
\cite{Chas} which indeed exclude {\it static} hairy configurations.
However, as we shall now show, time-dependent ({\it stationary})
resonances do exist!

In order to solve the resonance condition, Eq. (\ref{Eq16}), it is
instructive the introduce the dimensionless variable
\begin{equation}\label{Eq19}
\epsilon\equiv\sqrt{M^2\mu^2-(m/2)^2}\  ,
\end{equation}
in terms of which $\kappa$ and $\beta$ can be expressed as
\begin{equation}\label{Eq20}
\kappa={{(m/2)^2-\epsilon^2}\over{\epsilon}}\ \ \ \text{and}\ \ \
\beta^2=(l+{1\over 2})^2-{3\over
2}m^2+\epsilon^2+\sum_{k=1}^{\infty}c_k\epsilon^{2k}\  .
\end{equation}
Substituting (\ref{Eq20}) into (\ref{Eq16}), one finds that the
resonance condition can be expressed as a polynomial equation for
the dimensionless variable $\epsilon$ \cite{Notec1}:
\begin{eqnarray}\label{Eq21}
m^4-4m^2(2n+1)\epsilon+16\Big[m^2-(l+{1\over 2})^2+(n+{1\over
2})^2\Big]\epsilon^2 \nonumber
\\+16(2n+1)\epsilon^3-16c_1\epsilon^4
-16\sum_{k=2}^{\infty}c_k\epsilon^{2k+2} =0\ .
\end{eqnarray}



To demonstrate the existence of a discrete and infinite family of
stationary resonances, we present in Table \ref{Table1} the
dimensionless quantity $M\mu$ for the fundamental family ($l=m=1$)
of resonances. We display results for various values of the
resonance parameter $n$ \cite{Notelight}. A qualitatively similar
behavior is observed for other families of resonances (characterized
by the two parameters $\{l,m\}$).

The infinite spectrum $\{M\mu(n)\}$ of field-masses satisfying the
resonance condition (\ref{Eq16}) is a decreasing function (for
$n\geq 1$) of the resonance number $n$. For $n\gg l$ one finds from
(\ref{Eq21})
\begin{equation}\label{Eq22}
M\mu={m\over 2}+{{m^3}\over{16n^2}}-{{m^3}\over{8n^3}}+O(n^{-4})\
\end{equation}
for the resonances \cite{Notelm}.


\begin{table}[htbp]
\centering
\begin{tabular}{|c|c|c|c|}
\hline
$n$ & ${M\mu}_{\text{resonance}}$ & $x_{\text{cloud}}$ \\
\hline
\ 0\ \ &\ 0.526\ \ &\ 8.557\\
\ 1\ \ &\ 0.510\ \ &\ 23.485\\
\ 2\ \ &\ 0.505\ \ &\ 46.398\\
\ 3\ \ &\ 0.503\ \ &\ 77.313\\
\hline
\end{tabular}
\caption{Stationary scalar resonances of a maximally rotating Kerr
black hole. We display the dimensionless product $M\mu$ for the
fundamental family of resonances ($l=m=1$) and for various values of
the resonance parameter $n$. Also shown are the effective heights
(in units of the black-hole radius), $x_{\text{cloud}}$, of the
scalar clouds above the central black hole. In general, the size of
a scalar cloud increases with increasing resonance number $n$. Note
that all resonances conform to the lower bound (\ref{Eq23}).}
\label{Table1}
\end{table}

{\bf IV. Effective heights of the scalar clouds.\ \ } We shall next
evaluate the effective heights of the stationary scalar ``clouds"
which surround the central rotating black hole. These clouds
correspond to the family of wave-functions (\ref{Eq17}) satisfying
the resonance condition (\ref{Eq16}).

It is worth mentioning that a nice `no short hair theorem' was
proved in \cite{Nun,NoteHod11} for {\it static} and {\it
spherically} symmetric hairy black-hole configurations. According to
this theorem, the ``hairosphere" [the region where the non-trivial
(non-asymptotic) behavior of the black-hole hair is present] must
extend beyond ${3\over 2}$ the horizon radius. This translates into
the lower bound
\begin{equation}\label{Eq23}
x_{\text{hair}}\geq {1\over 2}\  ,
\end{equation}
where $x$ is the dimensionless height defined in (\ref{Eq7}). A
rough estimate of the size of our stationary scalar configurations
can be obtained by defining their effective radii as the radii at
which the quantity $4\pi r^2|\Psi|^2$ attains its global maximum.
Taking cognizance of Eq. (\ref{Eq17}) with $L^{(2\beta)}_0(z)\equiv
1$, one finds
\begin{equation}\label{Eq24}
x^{(0)}_{\text{cloud}}={{2\beta+1}\over{2\epsilon}}
\end{equation}
for the ground-state resonances with $n=0$ \cite{Noteheig}. For
higher $n$ values one generally obtains larger scalar clouds with
the approximated relation $x^{(n)}_{\text{cloud}}\simeq
{{2\beta+1+2n}\over{2\epsilon}}$, see Eq. (\ref{Eq17}).

The effective heights of the fundamental scalar clouds above the
rotating central black hole are given in Table \ref{Table1}, from
which we learn that $\{x_{\text{cloud}}(n)\}$ are always larger than
${1\over 2}$. For $n\gg l$ one finds from (\ref{Eq22})
$x^{(n)}_{\text{cloud}}\simeq 4n^2/m^2\gg 1$. It is remarkable that
our stationary scalar clouds respect the lower bound (\ref{Eq23})
despite the fact that they do {\it not} satisfy the conditions of
the original theorem \cite{Nun}: they are neither static nor
spherically symmetric.


{\bf V. Summary and discussion.\ \ } In this Letter we have analyzed
the dynamics of a massive scalar field in an astrophysically
realistic ({\it rotating}) Kerr black-hole spacetime. In particular,
we have proved the existence of a discrete and infinite family of
resonances describing stationary ({\it non}-decaying) scalar
configurations surrounding maximally rotating black holes. The
effective heights of these scalar ``clouds" above the central black
hole were determined analytically -- it was shown that
these non-static
and non-spherically symmetric configurations conform to the lower
bound (\ref{Eq23}) originally derived in \cite{Nun} for static and
spherically symmetric hairy black-hole configurations. Thus, our
analysis provides the first direct evidence for the general validity
of the bound (\ref{Eq23}).

Our results support the possible existence of stationary scalar
field dark matter distributions surrounding astrophysically
realistic (rotating) black holes. Moreover, while former studies
\cite{Barr,Det} focused on the regime $M\mu\ll 1$, our analysis
opens the possibility for the existence of such black hole--scalar
field configurations in the new regime $M\mu>{1/2}$. Our analytical
findings for maximally rotating black holes are in accord with the
numerical work of \cite{Dolan} who discussed slowly-decaying bound
states of non-extremal black holes.

Probably the most interesting question is the following one: what is
the {\it final} end-state of gravitational collapse? The no-hair
conjecture \cite{Whee} asserts that the final outcome of rotating
gravitational collapse is a bald Kerr black hole. However, our
results suggest the following alternative scenario: during the early
stages of the evolution some of the fields would indeed be radiated
away to infinity while some would be swallowed by the newly-born
black hole (see \cite{Barr} for the spherically symmetric case).
However, if the initial matter distribution contains massive scalar
fields, then those fields which satisfy the resonance condition
(\ref{Eq16}) may actually survive as {\it infinitely} long-lived
(stationary) resonances outside the black hole. Thus, the final
configuration is expected to be a rotating black hole surrounded by
stationary scalar clouds. It would be interesting to verify this
prediction using numerical simulations as done in \cite{Barr} for
the spherically symmetric (non-rotating) case.

\bigskip
\noindent
{\bf ACKNOWLEDGMENTS}
\bigskip

This research is supported by the Carmel Science Foundation. I thank
Yael Oren, Arbel M. Ongo and Ayelet B. Lata for helpful discussions.



\begin{thebibliography}{99}
\bibitem{Notenonlin} It should be emphasized that self-interaction effects tend to stabilize hairy
black-hole configurations \cite{Nun,Hod11} -- we therefore expect
our results to hold true in the non-linear regime as well.
\bibitem{Notelight} The resonances are characterized by small
values of $\epsilon$, in which case Eq. (\ref{Eq21}) yields
$M\mu(n)\simeq{1\over
2}+\big[{{2n+1-\sqrt{5}}\over{8[n(n+1)-1]}}\big]^2$. This analytical
expression agrees extremely well with the numerical data presented
in Table \ref{Table1}.
\bibitem{Notelm} Note that the regularity condition (\ref{Eq14}) restricts the
allowed values of the azimuthal harmonic index:
$m\leq\sqrt{{2}l(l+1)/3}$ for $n\gg l$ [see Eq. (\ref{Eq20})].
\bibitem{Noteheig} A lower bound on the effective heights (\ref{Eq24}) of the
stationary scalar clouds can be obtained from the couple of
inequalities (\ref{Eq14}) and (\ref{Eq18}):
$x^{(0)}_{\text{max}}>{{\sqrt{2}}\over{M\mu}}$. [Here we have used
the inequality $\epsilon<M\mu/\sqrt{2}$ which follows from
(\ref{Eq18}) and (\ref{Eq19}).]
\end{thebibliography}
\end{document}